\documentclass[12pt,a4paper]{article}
\usepackage{dfttrob}  
\usepackage{amsmath,amssymb}
\usepackage{graphicx}

\dfttnum{March 2003} 

\newcommand{\ee}{$e^+e^-$}               

\newcommand{\pt}{p_T}
\def\avg#1{\langle #1 \rangle}  
\def\nbar{\bar n}            

\begin{document}

\title{Schwinger Model and String Percolation\\ in Hadron--Hadron and
	Heavy Ion Collisions}
\author{J. Dias de Deus\\[-0.05cm]
 \textit{\small Departamento de F\'{\i}sica and CENTRA, Insituto Superior
	T\'ecnico}\\[-0.05cm]
 \textit{\small  Av. Rovisco Pais, 1049-001 Lisboa, Portugal}\\[0.2cm]
E.G. Ferreiro, C. Pajares\\[-0.05cm]
 \textit{\small Departamento de F\'{\i}sica de Part\'{\i}culas,}\\[-0.05cm]
 \textit{\small Universidade de Santiago de Compostela, 15706 Santiago de
	Compostela, Spain}\\[0.2cm]
R. Ugoccioni\\[-0.05cm]
 \textit{\small Dipartimento di Fisica Teorica and I.N.F.N - 
   Sezione di Torino}\\[-0.05cm]
 \textit{\small Via P. Giuria 1, 10125 Torino, Italy}}
\maketitle

\begin{abstract}
In the framework of the Schwinger Model for percolating strings we
establish a general relation between multiplicity and transverse
momentum square distributions in hadron--hadron and heavy ion
collisions. Some of our results agree with the Colour Glass Condensate
model. 
\end{abstract}

Experimental data from RHIC (Relativistic Heavy Ion Collider) show
very interesting features concerning particle rapidity densities and
transverse momentum, $\pt$, distributions 
\cite{Ullrich:2002tq+Bazilevsky:2002fz,Adcox:2001jp}.
They exclude high particle densities, expected in naive
multicollision models \cite{Braun:Feta+Braun:2000hd}, 
as well as fast growing values for
$\avg{\pt}$ as a function of energy, expected in naive perturbative
QCD models \cite{pQCD}.
Physics seems to remain classical and, essentially, non-perturbative.

Multiparticle production is frequently described as resulting from
multiple collisions at the parton level and, in the case of
nucleus--nucleus collisions, also at nucleon level, with formation of
colour strings stretched between the projectile and the target, which
decay into other strings that subsequently hadronize into the observed
hadrons \cite{DPM:1+SFM:1+SFM:2}.
There are long strings in rapidity, valence strings, associated to
valence quark (diquark) interactions, and short strings in rapidity,
centrally produced (sea strings) associated to 
interactions of sea partons, mostly gluons.
In a symmetrical AA collisions, with $N_A$ participants from each
nucleus, the number of valence strings equals the number of
participants, as in the wounded nucleon model 
\cite{WNM+Pajares:1985gw}, while the number of sea strings behaves
roughly as $N_s \approx N_A^{4/3}$ \cite{Armesto:SFM}, increasing with the
energy.

We shall adopt for the mechanism of particle production the Schwinger
model mechanism as developed in \cite{SchwingerModel,ColourRopes}.
In particular, the particle density and transverse momentum square
will be considered proportional to the field (and the charge) carried
by the string.

In multicollision models, many strings are produced, the number
increasing with energy, atomic mass and centrality. If the strings are
identical and independent, and approximately align with the collision
axis, we have, for the rapidity particle density, $d n/d y$, and for
the average of the square of the transverse momentum, $\avg{\pt^2}$,
\begin{equation}
	\frac{d n}{d y} = N_s \nbar_1 ,    \label{eq:1}
\end{equation}
\begin{equation}
	\avg{\pt^2} = \overline{p_1^2} ,         \label{eq:2}
\end{equation}
where $N_s$ is the number of strings, $\nbar_1$ is the single string
particle density and $\overline{p_1^2}$ the average transverse momentum
squared of the single string.

If the strings fuse in a rope \cite{ColourRopes}, 
the colour randomly grows as
$\sqrt{N_s}$ and we have 
\begin{equation}
	\frac{d n}{d y} = \frac{1}{\sqrt{N_s}} N_s \nbar_1 , \label{eq:3}
\end{equation}
\begin{equation}
	\avg{\pt^2} = \overline{p_1^2}\, \sqrt{N_s} .             \label{eq:4}
\end{equation}

In the situation of a hadron--hadron or nucleus--nucleus central
collision, the strings overlap in the impact parameter plane and the
problem becomes similar to a 2-dimensional continuum percolation
problem \cite{Pajares:perc+NardiSatz}.
If the strings are randomly distributed in the impact parameter plane
then, in the thermodynamical approximation \cite{Braun:2001us}, 
the overlapping colour reducing factor is given by
\begin{equation}
	F(\eta) = \sqrt{ \frac{ 1-e^{-\eta} }{ \eta } }   ,     \label{eq:5}
\end{equation}
where $\eta$ is the transverse density percolation parameter,
\begin{equation}
	\eta \equiv \left( \frac{r_s}{R} \right)^2 N_s ,      \label{eq:6}
\end{equation}
where $\pi r_s^2$ is the string transverse area and $\pi R^2$ the
interaction transverse area. We thus have
\begin{equation}
	\frac{d n}{d y} = F(\eta) N_s \nbar_1  ,                \label{eq:7}
\end{equation}
\begin{equation}
	\avg{\pt^2} = \frac{1}{ F(\eta) } \overline{p_1^2} .          \label{eq:8}
\end{equation}
Equations similar to (\ref{eq:7}) and (\ref{eq:8}) were written 
in \cite{Braun:2001us}.
As with $\eta \to 0$ (low density limit) $F(\eta)\to 0$ and with
$\eta\to\infty$ (high density limit) $F(\eta)\to1/\sqrt{\eta}$, the
behaviour of relations (\ref{eq:1}) and (\ref{eq:2}), 
and (\ref{eq:3}) and (\ref{eq:4}) is
recovered from (\ref{eq:7}) and (\ref{eq:8}).

We shall now discuss the consequences of (\ref{eq:7}) and
(\ref{eq:8}).  Two straightforward results follow:

\medskip\noindent i)
\emph{slow increase of particle density with energy and saturation of the
normalised particle densities as $N_s$ increases}

As the number of strings, $N_s$, increases with energy, at large
energy $\eta$ also increases and 
\begin{equation}
	F(\eta) \approx \frac1{\sqrt{\eta}} ,                     \label{eq:9}
\end{equation}
which means, (\ref{eq:7}),
\begin{equation}
	\frac{d n}{d y} \approx \left( \frac R {r_s} \right) N_s^{1/2} \nbar_1 .
	                                                      \label{eq:10}
\end{equation}
Instead of growing with $N_s$, as one should have naively expected
with independent strings, (\ref{eq:1}), the density grows more slowly, as
$N_s^{1/2}$.

On the other hand, as
\begin{equation}
	N_s \approx N_A^{4/3} \quad,\qquad R\approx R_1 N_A^{1/3} , \label{eq:11}
\end{equation}
where $R_1$ is a quantity of the order of the nucleon radius,
\begin{equation}
	\frac{1}{N_A} \frac{d n}{d y} \approx 
	    \left( \frac{R_1}{r_s} \right) \nbar_1              \label{eq:12}
\end{equation}
tends to saturate as $N_A$ increase. Both behaviours 
(\ref{eq:7}) and (\ref{eq:12}) were confirmed by data 
\cite{Ullrich:2002tq+Bazilevsky:2002fz}.

The saturation, in our framework, is a consequence of string
percolation \cite{RU:dNdeta+RU:dNdeta:2}. 
At the level of QCD it can be seen as resulting from
low-$x$ parton saturation in the colliding nuclei \cite{saturation}.

\medskip\noindent ii)
\emph{a universal relation between $dn/dy$ and $\avg{\pt}$}

For large density, Eqs.~(\ref{eq:7}) and (\ref{eq:8}) become
\begin{equation}
	\frac{d n}{d y} =  \left( \frac R {r_s} \right) N_s^{1/2} \nbar_1  ,
                                                \label{eq:13}
\end{equation}
\begin{equation}
	\avg{\pt^2} =  \left( \frac {r_s} R \right) N_s^{1/2} \overline{p_1^2} ,
                                            \label{eq:14}
\end{equation}
and, eliminating $N_s^{1/2}$,
\begin{equation}
	\sqrt{\avg{\pt^2}} = c \sqrt{ \frac{1}{N_A^{2/3}} \frac{dn}{dy} } ,
	                                              \label{eq:15}
\end{equation}
with
\begin{equation}
	c \equiv \left( \frac {r_s} {R_1} \right) 
	           \left( \frac{\overline{p_1^2}}{\nbar_1} \right)^{1/2} .
                                                 \label{eq:16}
\end{equation}
A relation of this type,
\begin{equation}
	\sqrt{\avg{\pt^2}} \approx \sqrt{ \frac{1}{N_A^{2/3}} \frac{dn}{dy}
	}                                             \label{eq:17}
\end{equation}
was obtained, in the framework of the Colour Glass Condensate (CGC)
model \cite{ColourGlass}, in \cite{McLerran:2001cv+Schaffner-Bielich:2001qj}.
Our formula (\ref{eq:14}) includes not only the functional dependence, but,
as well, the proportionality factor $c$. 

We can make an order of magnitude estimate of the proportionality
factor $c$. 
In the dual string model $r_s\approx$ 0.2~fm 
\cite{Pajares:perc+NardiSatz,Satz:98}, $R_1$ should be of
the order of the proton radius ($\approx$ 1~fm) and for the 
string charged particle production parameters one has $\bar p_1
\approx 0.3$ and $\nbar_1 \approx 0.7$, as observed from low energy
data \cite{giacomelli}, and $(\overline{p_1^2}/\nbar_1)^{1/2}\approx 0.35$.
The proportionality factor is then $\approx$ 0.07 to be compared with
0.0348 for pions and 0.100 for kaons 
\cite{McLerran:2001cv+Schaffner-Bielich:2001qj}.
In the comparison with data we shall identify $\sqrt{\avg{\pt^2}}$
with $\avg{\pt}$ and $\sqrt{\overline{p_1^2}}$ with $\bar p_1$
(this overestimates the average values of $\avg{\pt}$ and $\bar p_1$).

We have just considered the high $\eta$ limit.
In the low density end, which means low energy and peripheral
collisions, we have just valence strings and $\avg{\pt} \to
\bar p_1\approx 0.3~\text{GeV}$.
This is, in practice, the value of $\avg{\pt}$ in pp collisions at low
($\sqrt{s} \lesssim 10~\text{GeV}$) energies.

\begin{figure}
  \begin{center}
  \mbox{\includegraphics[width=0.7\textwidth]{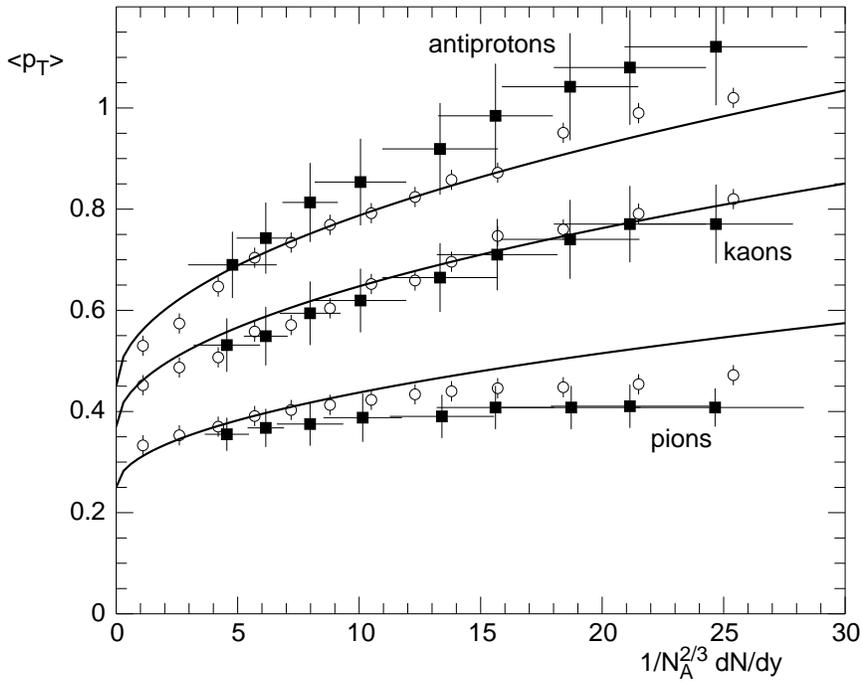}}
  \end{center}
  \caption[pt vs dn/dy]{$\avg{\pt}$ vs multiplicity density in p\=p collisions
	(where $N_A=1$) at 1800 GeV \cite{Alexopoulos:1993wt} (open circles)
	and in central Au+Au collisions at 200 $A$GeV \cite{VanBuren:2002} (filled 
	squares). Solid lines represent Eq.~(\ref{eq:18}) with $\bar p_1$
	adjusted separately to each species.}\label{fig:1}
  \end{figure}

By putting these two limits together, we arrive at the formula
obtained in \cite{McLerran:2001cv+Schaffner-Bielich:2001qj}, 
but now with all the parameters theoretically
constrained:
\begin{equation}
	\avg{\pt} = \bar p_1 \left( 1 + \frac{r_s}{R}
	         \frac{1}{\nbar_1^{1/2}} \sqrt{
						\frac{1}{N_A^{2/3}} \frac{dn}{dy} } \right) .  
                                                    \label{eq:18}
\end{equation}
In Fig.~\ref{fig:1} we compare Eq.~(\ref{eq:18}) with data.
The agreement is not perfect, but there is an indication that some
truth exists in CGC and string percolation models.

In the next step we make an attempt to generalise our results and to
relate the (normalised) transverse momentum distribution  $f(\pt^2)$
to the multiplicity distribution $P(n)$, in hadron--hadron and
nucleus--nucleus collisions.

We work in the large $\eta$ limit and start by changing the notation,
and write
\begin{equation}
	N = \alpha N_s^{1/2} ,                             \label{eq:19}
\end{equation}
with
\begin{equation}
	\alpha \equiv R/r_s ,                              \label{eq:20}
\end{equation}
such that $N$ has the meaning of the number of effective strings
(mostly sea strings or ropes). If $n$ particles are produced
\begin{equation}
	n = N n_1 ,                                        \label{eq:21}
\end{equation}
see also Eq.~(\ref{eq:13}). This effective number $N$ takes into account
percolation effects in the sum of colours of the $N_s$ individual
strings. 

Let $P(N)$ be the probability of producing $N$ effective identical
strings and $p(n_i)$ the probability of producing $n_i$ particles from
the $i$-th string. We then have 
\begin{equation}
	P(n) = \int  P(N) \prod_{i=1}^N p(n_i) d n_i 
	       \delta\left( n - \sum_{i=1}^N n_i\right) d N   \label{eq:22}
\end{equation}
In (\ref{eq:22}), as the colour percolation effects were absorbed in $N$, we
treated the effective strings as independent (see \cite{components}).

Regarding transverse momentum distributions, the natural
generalisation for (\ref{eq:14}) is to write
\begin{equation}
	\pt^2 = \frac{N}{\alpha^2} p_1^2 ,            \label{eq:23}
\end{equation}
and for the distribution itself
\begin{equation}
	F(\pt^2) = \int  P(N) f(p_1^2) \delta\left( \pt^2 - 
	             \frac{N}{\alpha^2} p_1^2 \right) dp_1^2 dN . \label{eq:24}
\end{equation}
In this case, for a given $F(\pt^2)$ contribute all effective strings
with $f(p_1^2)$, such that $p_1^2$ satisfies (\ref{eq:23}).
As all strings are assumed equal, $f(p_1^2)$ is representative of any
string. 

In order to construct $P(n)$ and $F(\pt^2)$, one of course needs the
elementary string distributions $p(n_1)$ and $f(p_1^2)$ and the
distribution $P(N)$ of effective strings.
Concerning the $p(n_1)$ distribution, it should be Poisson or close to
Poisson type (as seen in \ee\ at low energy \cite{HRS:1}). 
The $\pt^2$ distribution in the Schwinger model is an exponential in
$-\pt^2$. 
The $P(N)$ distribution contains the nucleonic and the partonic
structure of the colliding particles and the combinatorial factors of
Glauber-Gribov calculus; its shape is investigated in \cite{Pajares:II}.

Our objective here is not to solve Eqs.~(\ref{eq:22}) and (\ref{eq:24}), 
but simply to try to relate $P(n)$ to $F(\pt^2)$.
In view of that, let us proceed by calculating the $\avg{n^q}$ and
$\avg{\pt^{2q}}$ moments of the distributions (\ref{eq:22}) and (\ref{eq:24}),
respectively.
The calculations are straightforward, but lengthy in the case of
multiplicities (see, for example, \cite{components}).
In this case, to simplify, we shall assume
\begin{equation}
	p(n_1) = \delta(n_1 - \nbar_1) .              \label{eq:25}
\end{equation}
It has been shown, sometime ago, that this approximation in
hadron--hadron and nucleus--nucleus collisions is very reasonable
\cite{components}. 

We have then for the moments:
\begin{equation}
	\avg{n^q} = \avg{N^q} \nbar_1^q .             \label{eq:26}
\end{equation}

It is clear, because of (\ref{eq:25}), that all fluctuations come from
fluctuations in the number of effective strings. For $\pt$
distribution 
\begin{equation}
	\avg{\pt^{2q}} = \frac{\avg{N^q}}{\alpha^{2q}} \bar p_1^{2q} . \label{eq:27}
\end{equation}
Eqs.~(\ref{eq:26}) and (\ref{eq:27}) are the natural generalisation 
of (\ref{eq:13}) and (\ref{eq:14}).
As before, the moments of the effective string distribution can be
eliminated by dividing (\ref{eq:27}) by (\ref{eq:26})
and a relation between $\avg{n^q}$ and
$\avg{\pt^{2q}}$ established.
But one can now do better and eliminate the strongly model-dependent
parameter $\alpha=R/r_s$, eq.~(\ref{eq:20}).
If one writes the KNO moments
\begin{equation}
	C_q^X \equiv \frac{ \avg{X^q} }{ \avg{X}^q } \quad,\qquad
	      q = 1, 2, \dots ,                             \label{eq:28}
\end{equation}
the parameter $\alpha$ disappears.
By using a capital $C$ for final distributions KNO moments and a small
$c$ for single string distributions KNO moments, our final result can
be written as
\begin{equation}
	\frac{ C_q^n}{c_q^n} = \frac{ C_q^{\pt^2} }{ c_q^{\pt^2} } .
                                                    \label{eq:29}
\end{equation}
This equation, as mentioned before, is strictly correct only for
$c_q^n=1$. 

It is not easy to check Eq.~(\ref{eq:29}) accurately, 
as most experiments can only measure $\pt > \avg{\pt}$, but 
one can nonetheless attempt a somewhat rough comparison.
In the Schwinger model the $\pt$ distribution is Gaussian, which means 
$c_q^{\pt^2} = q!$. 
If the final $\pt$ distribution is also a Gaussian, then one obtains
$C_q^n = 1$, which is not a good approximation.
If the final $\pt$ distribution is an exponential, which is closer to
reality \cite{Abe:1988yu}, then $C_q^{\pt^2}=(2q+1)!/(3!)^2$, 
and we obtain, for instance, $C_2^n = 5!/(3!)^2 2! \approx 1.66$.
This is to be compared with the experimental value $C_2^n \approx 1.3$
at $\sqrt{s} = 200$ GeV \cite{UA5:3}. 

Finally, the main point we want to make with (\ref{eq:29}) is that
multiplicity and transverse momentum distributions are deeply
related: one should remember that, in general, from the $C_q^x$ moments
one can construct the distribution in KNO form,
$\avg{x} P(x/\avg{x})$.

\section*{Acknowledgements}
Two of us (E.G.F. and C.P.) thank the financial support of the CICYT of
Spain and E.U. through the contract FPA2002-01161.


\end{document}